\documentstyle[preprint,aps,floats,epsfig]{revtex}
\tightenlines

\def\Journal#1#2#3#4{{#1} {#2} (#4) #3}
\def\EPJC{Eur. Phys. J. C}
\def\NP{Nucl. Phys.}
\def\PLB{Phys. Lett. B}
\def\PRC{Phys. Rev. C}
\def\PRD{Phys. Rev. D}
\def\PRL{Phys. Rev. Lett.}
\def\PR{Phys. Rept.}
\def\RPP{Rept. Prog. Phys.}

\def\ZPC{Z. Phys. C}

\begin{document}
\title{$\Upsilon$ absorption in hadronic matter}
\bigskip
\author{Ziwei Lin and C. M. Ko}
\address{Cyclotron Institute and Physics Department, Texas A\&M University,
College Station, Texas 77843-3366}
\maketitle

\begin{abstract}
The cross sections of $\Upsilon$ absorption by $\pi$ and $\rho$ mesons 
are evaluated in a meson-exchange model. 
Including form factors with a cutoff parameter of $1$ or $2$ GeV, 
we find that due to the large threshold of these reactions the thermal average 
of their cross sections is only about $0.2$ mb at a temperature of $150$ MeV.
Our results thus suggest that the absorption of directly produced $\Upsilon$ 
by hadronic comovers in high energy heavy ion collisions is unimportant.

\medskip
\noindent PACS number(s): 25.75.-q, 13.75.Lb, 14.40.Gx, 14.40.Nd
\end{abstract}

\section{Introduction}

Recent experiments \cite{na50} at the CERN SPS have shown an anomalously 
large suppression of $J/\psi$ production in central Pb+Pb collisions.
Following the original idea of Matsui and Satz \cite{matsui}
that $J/\psi$ would be dissociated in a quark-gluon plasma due to
color screening, the observed $J/\psi$ suppression has been suggested 
as an evidence for the formation of the
quark-gluon plasma in these collisions \cite{blaiz,wong,khar}. 
On the other hand, it has also been shown that 
$J/\psi$ absorption by comoving hadrons in the dense matter 
is important if the cross sections are taken to be a few mb 
\cite{cassing,capella,kahana,gale,spieles,sa}.
Although these cross sections are much larger than those predicted
in earlier theoretical studies based on either the perturbative
QCD \cite{satz} or a simple hadronic Lagrangian \cite{bm},
they are consistent with recent studies using the
quark-exchange model \cite{blaschke,swanson} 
or a more general hadronic Lagrangian \cite{haglin,jpsih}. 

Since bottomonium states in a quark-gluon
plasma are also sensitive to the color screening effect \cite{matsui,review}, 
the study of $\Upsilon$ suppression in high energy heavy ion collisions 
can be used as a signature for the quark-gluon plasma as well.
Because of its larger binding energy than that of $J/\psi$, 
the critical energy density at which an $\Upsilon$ is dissociated in 
the quark-gluon plasma is also higher \cite{karsch}. One thus expects
to see the effects of the quark-gluon plasma on the production of $\Upsilon$ 
only in ultra-relativistic heavy ion collisions 
such as at the BNL RHIC and the CERN LHC.
As in the case of $J/\psi$, one needs to understand the
effects of $\Upsilon$ absorption in hadronic matter in
order to use its suppression as a signal for the quark-gluon plasma 
in heavy ion collisions.  In this paper, we shall study the
$\Upsilon$ absorption cross sections by $\pi$ and $\rho$ mesons,
which are the dominant hadrons in ultra-relativistic heavy ion collisions.

This paper is organized as follows. In Sec.~\ref{sec_lagn}, we introduce 
the hadronic Lagrangian and derive the relevant 
interaction Lagrangians between $\Upsilon$ and other hadrons.
The cross sections for 
$\Upsilon$ absorption by $\pi$ and $\rho$ mesons are then evaluated in 
Sec.~\ref{sec_ampl}, and the numerical results are given  
in Sec.~\ref{sec_num}.
In Sec.~\ref{sec_j}, we show the relation between the cross sections 
for $\Upsilon$ absorption and those for $J/\psi$ absorption.
Finally, discussions and a summary are given in Sec.~\ref{sec_sum}.

\section{The hadronic Lagrangian}
\label{sec_lagn}

We start from the following SU(5) symmetric free Lagrangian 
for pseudoscalar and vector mesons:
\begin{eqnarray}
{\cal L}_0= {\rm Tr} \left ( \partial_\mu P^\dagger \partial^\mu P \right )
-\frac{1}{2} {\rm Tr} \left ( F^\dagger_{\mu \nu} F^{\mu \nu} \right )~,
\label{lagn0}
\end{eqnarray}
where $F_{\mu \nu}=\partial_\mu V_\nu-\partial_\nu V_\mu$, 
and $P$ and $V$ denote, respectively, the $5\times 5$ 
pseudoscalar and vector meson matrices in SU(5): 
{\footnotesize
\begin{eqnarray} P&=&
\frac{1}{\sqrt 2}\left (
\begin{array}{ccccc}
\frac{\pi^0}{\sqrt 2}+\frac{\eta}{\sqrt 6}
+\frac{\eta_c}{\sqrt {12}}+\frac{\eta_b}{\sqrt {20}}
& \pi^+ & K^+ & \bar {D^0} & B^+ \\
\pi^- & -\frac{\pi^0}{\sqrt 2}+\frac{\eta}{\sqrt 6}
+\frac{\eta_c}{\sqrt {12}}+\frac{\eta_b}{\sqrt {20}}
& K^0 & D^- & B^0 \\
K^- & \bar {K^0} & -{\frac{2\eta}{\sqrt 6}}
+\frac{\eta_c}{\sqrt {12}}+\frac{\eta_b}{\sqrt {20}}
& D_s^- & B_s^0 \\
D^0 & D^+ & D_s^+ & 
-\frac{3\eta_c}{\sqrt {12}}+\frac{\eta_b}{\sqrt {20}}
& B_c^+ \\
B^- & \bar {B^0} & \bar {B_s^0} & B_c^- & 
-\frac{2\eta_b}{\sqrt 5}
\end{array}
\right ) \;, \nonumber \\[2ex]
V&=&\frac{1}{\sqrt 2}\left (
\begin{array}{ccccc}
\frac{\rho^0}{\sqrt 2}+\frac{\omega}{\sqrt 6}
+\frac{J/\psi}{\sqrt {12}}+\frac{\Upsilon}{\sqrt {20}}
& \rho^+ & K^{*+} & \bar {D^{*0}} & B^{*+} \\
\rho^- & -\frac{\rho^0}{\sqrt 2}+\frac{\omega}{\sqrt 6} 
+\frac{J/\psi}{\sqrt {12}}+\frac{\Upsilon}{\sqrt {20}}
& K^{*0} & D^{*-} & B^{*0}\\
K^{*-} & \bar {K^{*0}} & 
-{\frac{2\omega}{\sqrt 6}}+\frac{J/\psi}{\sqrt {12}}
+\frac{\Upsilon}{\sqrt {20}} & D_s^{*-} & B_s^{*0}\\
D^{*0} & D^{*+} & D_s^{*+} & 
-\frac{3J/\psi}{\sqrt {12}}+\frac{\Upsilon}{\sqrt {20}} & B_c^{*+} \\
B^{*-} & \bar {B^{*0}} & \bar {B_s^{*0}} & B_c^{*-} & 
-\frac{2\Upsilon}{\sqrt 5}
\end{array}
\right ) \;. 
\label{pv}
\end{eqnarray}
}
Introducing the minimal substitution as in Refs. \cite{jpsih,dmeson},
\begin{eqnarray}
\partial_\mu P &\rightarrow& {\cal D}_\mu P= \partial_\mu P
-\frac{ig}{2} \left [V_\mu, P \right ]~, \label{ms1} \\
F_{\mu \nu} &\rightarrow& 
\partial_\mu V_\nu-\partial_\nu V_\mu -\frac{ig}{2} 
\left [ V_\mu, V_\nu \right ]~,
\label{ms2}
\end{eqnarray}
leads to the following interaction hadronic Lagrangian:
\begin{eqnarray}
{\cal L}&=& {\cal L}_0 + ig {\rm Tr} 
\left ( \partial^\mu P \left [P, V_\mu \right ] \right ) 
-\frac{g^2}{4} {\rm Tr} 
\left ( \left [ P, V_\mu \right ]^2 \right ) \nonumber \\
&+& ig {\rm Tr} \left ( \partial^\mu V^\nu \left [V_\mu, V_\nu \right ] 
\right ) 
+\frac{g^2}{8} {\rm Tr} \left ( \left [V_\mu, V_\nu \right ]^2 \right )~.
\label{lagn2}
\end{eqnarray}
Since the SU(5) symmetry is explicitly broken by hadron masses, mass terms
based on the experimentally determined values 
are added to the above hadronic Lagrangian. 

Expanding the Lagrangian in Eq. (\ref{lagn2}) with 
the pseudoscalar meson and vector meson matrices shown in Eq. (\ref{pv}),  
we obtain the following interaction Lagrangians that are relevant 
for the absorption of $\Upsilon$ by $\pi$ and $\rho$ mesons:
\begin{eqnarray}
{\cal L}_{\pi BB^*}&=&ig_{\pi BB^*}~ 
\bar {B^{* \mu}} \vec \tau \cdot \left ( B \partial_\mu \vec \pi - 
\partial_\mu B \vec \pi \right ) + {\rm H.c.}~ , 
\nonumber \\ 
{\cal L}_{\Upsilon BB}&=&ig_{\Upsilon BB}~ \Upsilon^\mu 
\left ( \bar B \partial_\mu B - \partial_\mu \bar B B \right ) ~ ,
\nonumber \\ 
{\cal L}_{\Upsilon B^*B^*}&=& ig_{\Upsilon B^*B^*}~
\left [ \Upsilon^\mu \left ( \partial_\mu \bar {B^{* \nu}} B^*_\nu 
-\bar {B^{* \nu}} \partial_\mu B^*_\nu \right )
+\left ( \partial_\mu \Upsilon^\nu \bar {B^*_\nu} 
-\Upsilon^\nu \partial_\mu \bar {B^*_\nu}
\right ) B^{* \mu} \right .\nonumber \\ 
&+&\left . \bar {B^{* \mu}} \left ( \Upsilon^\nu \partial_\mu B^*_\nu 
-\partial_\mu \Upsilon^\nu B^*_\nu \right ) \right ] ~ , 
\nonumber \\ 
{\cal L}_{\pi \Upsilon BB^*}&=&-g_{\pi \Upsilon BB^*}~
\Upsilon^\mu \left ( \bar {B^*_\mu} \vec \tau B 
+ \bar B \vec \tau B^*_\mu \right ) \cdot \vec \pi ~ , 
\nonumber \\ 
{\cal L}_{\rho BB}&=&ig_{\rho BB}~ \left ( \bar B \vec \tau \partial_\mu B
-\partial_\mu \bar B \vec \tau B \right ) \cdot \vec \rho^\mu ~ ,
\nonumber \\ 
{\cal L}_{\rho \Upsilon BB}&=&g_{\rho \Upsilon BB}~ 
\Upsilon^\mu \bar B \vec \tau B \cdot \vec {\rho_\mu} ~ ,
\nonumber \\ 
{\cal L}_{\rho B^*B^*}&=&ig_{\rho B^*B^*}~ \left [ 
\left ( \partial_\mu \bar {B^{* \nu}} \vec \tau B^*_\nu
-\bar {B^{* \nu}} \vec \tau \partial_\mu B^*_\nu \right ) \cdot \vec \rho^\mu
+\left ( \bar {B^{* \nu}} \vec \tau \cdot \partial_\mu \vec \rho_\nu 
-\partial_\mu \bar {B^{* \nu}} \vec \tau \cdot \vec \rho_\nu \right ) 
B^{* \mu} \right .\nonumber \\ 
&+& \left . \bar {B^{* \mu}} \left ( 
\vec \tau \cdot \vec \rho^\nu \partial_\mu B^*_\nu
-\vec \tau \cdot \partial_\mu \vec \rho^\nu B^*_\nu \right ) \right ]~ ,
\nonumber \\ 
{\cal L}_{\rho \Upsilon B^*B^*}&=&g_{\rho \Upsilon B^*B^*}~ 
\left ( \Upsilon^\nu \bar {B^*_\nu} \vec \tau B^*_\mu 
+\Upsilon^\nu \bar {B^*_\mu} \vec \tau B^*_\nu 
-2 \Upsilon_\mu \bar {B^{* \nu}}\vec \tau B^*_\nu 
\right ) \cdot \vec \rho^\mu ~ .
\label{cc}
\end{eqnarray}
In the above, 
$B$ and $B^*$ denote, respectively, the pseudoscalar and 
vector bottom meson doublets, e.g., $B=(B^+, B^0)^{\rm T}$.  

\section{$\Upsilon$ absorption cross sections}
\label{sec_ampl}

\begin{figure}[ht]
\centerline{\epsfig{file=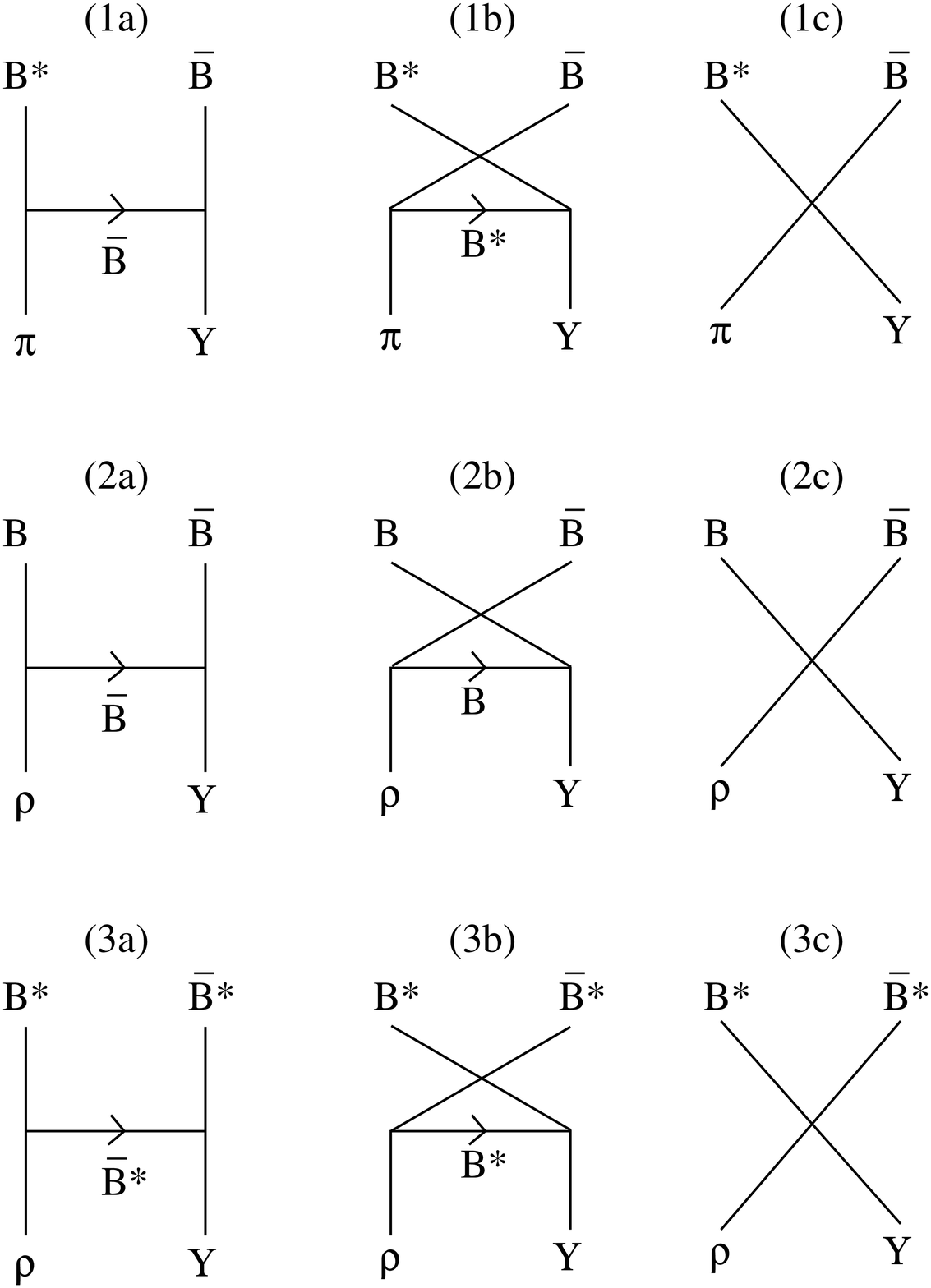,height=3in,width=3in,angle=0}}
\vspace{1cm}
\caption{Diagrams for $\Upsilon$ absorption processes:  
(1) $\pi \Upsilon \rightarrow B^* \bar B$,
(2) $\rho \Upsilon \rightarrow B \bar B$, 
and (3) $\rho \Upsilon \rightarrow B^* \bar {B^*}$. 
Diagrams for the process $\pi \Upsilon \rightarrow B \bar {B^*}$ are similar 
to (1) but with each particle replaced by its anti-particle.
}
\label{diagrams}
\end{figure}

The above hadronic Lagrangians allow us to study the following processes
for $\Upsilon$ absorption by $\pi$ and $\rho$ mesons:
\begin{eqnarray}
\pi \Upsilon \rightarrow B^* \bar B, 
~ \pi \Upsilon \rightarrow B \bar {B^*},
~\rho \Upsilon \rightarrow B \bar B, 
~ \rho \Upsilon \rightarrow B^* \bar {B^*}. 
\label{proc}
\end{eqnarray}
Corresponding diagrams for these processes
are shown in Fig.~\ref{diagrams}.

The total amplitude for the first process, 
$\pi \Upsilon \rightarrow B^* \bar B$, 
without isospin factors and before averaging (summing) 
over initial (final) spins, is given by
\begin{eqnarray}
{\cal M}_1 
=\left ( \sum_{i=a,b,c} {\cal M}_{1 i}^{\nu \lambda} \right )
\epsilon_{2 \nu} \epsilon_{3 \lambda}
\equiv {\cal M}_1^{\nu \lambda} ~\epsilon_{2 \nu} \epsilon_{3 \lambda},  
\end{eqnarray}
where the partial amplitudes for diagrams (1a), (1b), and (1c) are,
respectively, 
\begin{eqnarray}
{\cal M}_{1a}^{\nu \lambda}&=& -g_{\pi B B^*} g_{\Upsilon B B}~
(-2p_1+p_3)^\lambda \left (\frac{1}{t-m_B^2} \right ) 
(p_1-p_3+p_4)^\nu, \nonumber \\
{\cal M}_{1b}^{\nu \lambda}&=& g_{\pi B B^*} g_{\Upsilon B^* B^*}~
(-p_1-p_4)^\alpha \left ( \frac{1}{u-m_{B^*}^2} \right )
\left [ g_{\alpha \beta}-\frac{(p_1-p_4)_\alpha (p_1-p_4)_\beta}{m_{B^*}^2}
\right ] \nonumber \\
&\times& \left [ (-p_2-p_3)^\beta g^{\nu \lambda}
+(-p_1+p_2+p_4)^\lambda g^{\beta \nu}
+(p_1+p_3-p_4)^\nu g^{\beta \lambda} \right ] , \nonumber \\
{\cal M}_{1c}^{\nu \lambda}&=& -g_{\pi \Upsilon B B^*}~ g^{\nu \lambda}. 
\label{m_i}
\end{eqnarray}
In the above, $p_j$ denotes the momentum of particle $j$.
We use the notation that particles $1$ and $2$ represent 
the initial-state mesons while particles $3$ and $4$ 
represent the final-state mesons on the left and right side of the diagrams 
shown in Fig.~\ref{diagrams}, respectively.
The indices $\mu, \nu, \lambda$, and $\omega$ 
denote the polarization components of external vector mesons while
the indices $\alpha$ and $\beta$ denote those of the exchanged vector mesons. 
 
Similarly, the partial amplitudes for the second process, 
$\rho \Upsilon \rightarrow B \bar B$, are given by
\begin{eqnarray}
{\cal M}_{2a}^{\mu \nu}&=& -g_{\rho B B} g_{\Upsilon B B}~
(p_1-2 p_3)^\mu \left ( \frac{1}{t-m_B^2} \right ) 
(p_1-p_3+p_4)^\nu, \nonumber \\
{\cal M}_{2b}^{\mu \nu}&=& -g_{\rho B B} g_{\Upsilon B B}~
(-p_1+2 p_4)^\mu\left ( \frac{1}{u-m_B^2}\right ) (-p_1-p_3+p_4)^\nu, 
\nonumber \\
{\cal M}_{2c}^{\mu \nu}&=& g_{\rho \Upsilon B B}~ g^{\mu \nu},
\end{eqnarray}
and those for the third process, 
$\rho \Upsilon \rightarrow B^* \bar {B^*}$, are given by
\begin{eqnarray}
{\cal M}_{3a}^{\mu \nu \lambda \omega}&=& 
g_{\rho B^* B^*} g_{\Upsilon B^* B^*}~
\left [ (-p_1-p_3)^\alpha g^{\mu \lambda} +2 ~p_1^\lambda g^{\alpha \mu}
+2 p_3^\mu g^{\alpha \lambda} \right ] \left ( \frac{1}{t-m_{B^*}^2} \right )
\nonumber \\
&\times& \left [ g_{\alpha \beta}-\frac{(p_1-p_3)_\alpha (p_1-p_3)_\beta}
{m_{B^*}^2} \right ] \left [ -2 p_2^\omega g^{\beta \nu}
+(p_2+p_4)^\beta g^{\nu \omega}
-2 p_4^\nu g^{\beta \omega} \right ] , \nonumber \\
{\cal M}_{3b}^{\mu \nu \lambda \omega}&=& 
g_{\rho B^* B^*} g_{\Upsilon B^* B^*}~
\left [ -2 p_1^\omega g^{\alpha \mu} + (p_1+p_4)^\alpha g^{\mu \omega}
-2 p_4^\mu g^{\alpha \omega} \right ] \left ( \frac{1}{u-m_{B^*}^2} \right )
\nonumber \\
&\times& \left [ g_{\alpha \beta}-\frac{(p_1-p_4)_\alpha (p_1-p_4)_\beta}
{m_{B^*}^2} \right ] \left [ (-p_2-p_3)^\beta g^{\nu \lambda}
+2 p_2^\lambda g^{\beta \nu}+2 p_3^\nu g^{\beta \lambda}\right ], \nonumber \\
{\cal M}_{3c}^{\mu \nu \lambda \omega}&=& g_{\rho \Upsilon B^* B^*}~
\left ( g^{\mu \lambda} g^{\nu \omega} + g^{\mu \omega} g^{\nu \lambda} 
-2 g^{\mu \nu} g^{\lambda \omega} \right ).
\label{m_f}
\end{eqnarray}

Since the interaction Lagrangian in Eq. (\ref{lagn2}) is
generated by the minimal substitution, which is equivalent
to treating vector mesons as gauge particles,  
the total scattering amplitude for each process should satisfy
the condition of current conservation 
in the limit of zero vector meson masses, 
degenerate pseudoscalar meson masses, and SU(5) coupling constants,
e.g., ${\cal M}_1^{\nu \lambda} p_{3 \lambda}=0$.
One can easily check that the amplitudes given in Eqs. (\ref{m_i})-(\ref{m_f}) 
all satisfy the current conservation condition.  

After averaging (summing) over initial (final) spins 
and including isospin factors, the cross sections are 
\begin{eqnarray}
\frac {d\sigma_1}{dt}&=& \frac {1}{96 \pi s p_{i,\rm cm}^2} 
{\cal M}_1^{\nu \lambda} {\cal M}_1^{*\nu^\prime \lambda^\prime}
\left ( g_{\nu \nu^\prime}-\frac{p_{2 \nu} p_{2 \nu^\prime}} {m_2^2} \right )
\left ( g_{\lambda \lambda^\prime}
-\frac{p_{3 \lambda} p_{3 \lambda^\prime}} {m_3^2} \right ), \label{jpion}\\
\frac {d\sigma_2}{dt}&=&\frac {1}{288 \pi s p_{i,\rm cm}^2}
{\cal M}_2^{\mu \nu} {\cal M}_2^{*\mu^\prime \nu^\prime}
\left ( g_{\mu \mu^\prime}-\frac{p_{1 \mu} p_{1 \mu^\prime}} {m_1^2} \right )
\left ( g_{\nu \nu^\prime}-\frac{p_{2 \nu} p_{2 \nu^\prime}} {m_2^2} \right ),
\\
\frac {d\sigma_3}{dt}&=&\frac {1}{288 \pi s p_{i,\rm cm}^2} 
{\cal M}_3^{\mu \nu \lambda \omega} 
{\cal M}_3^{*\mu^\prime \nu^\prime \lambda^\prime \omega^\prime}
\left ( g_{\mu \mu^\prime}-\frac{p_{1 \mu} p_{1 \mu^\prime}} {m_1^2} \right )
\left ( g_{\nu \nu^\prime}-\frac{p_{2 \nu} p_{2 \nu^\prime}} {m_2^2} \right ) 
\nonumber \\
&\times& \left ( g_{\lambda \lambda^\prime}
-\frac{p_{3 \lambda} p_{3 \lambda^\prime}} {m_3^2} \right )
\left ( g_{\omega \omega^\prime}
-\frac{p_{4 \omega} p_{4 \omega^\prime}} {m_4^2} \right ) ,
\label{m3}
\end{eqnarray}
with $s=(p_1+p_2)^2$, and $p_{i,\rm cm}$ denoting the momentum of 
each initial-state meson in the center-of-mass frame. 

With the exact SU(5) symmetry, 
the coupling constants in Eq. (\ref{cc})
can be related to the SU(5) universal coupling constant $g$ by
the following relations:
\begin{eqnarray}
&&g_{\pi BB^*}=g_{\rho BB}=g_{\rho B^* B^*}=\frac{g}{4}~, ~
g_{\Upsilon BB}=g_{\Upsilon B^* B^*}=\frac{5g}{4 \sqrt {10}}~, \nonumber \\
&&g_{\pi \Upsilon BB^*}=g_{\rho \Upsilon B^* B^*}=\frac{5g^2}{16 \sqrt {10}}~,~
g_{\rho \Upsilon B B}=\frac{5g^2}{8 \sqrt {10}}~.   
\label{su5}
\end{eqnarray}
These coupling constants can be further related to those involving
light and charm mesons, i.e.,
\begin{eqnarray}
g_{\rho \pi \pi}=2g_{\pi B B^*}=2g_{\pi D D^*}=
\sqrt {\frac {8}{5}} g_{\Upsilon B B}=\sqrt {\frac {3}{2}} g_{\psi DD}.  
\label{su5r}
\end{eqnarray}
Values of the light and charm meson coupling constants are known
\cite{jpsih}, and they are $g_{\rho \pi \pi}=6.1$, $g_{\pi D D^*}\simeq 4.4$, 
and $g_{\psi DD} \simeq 7.6$. 

The three-point coupling constants for bottom mesons 
can also be determined phenomenologically.
Using the vector meson dominance model as in Ref. \cite{jpsih}
for charm mesons, we obtain 
\begin{eqnarray}
g_{\rho BB}=g_{\rho B^* B^*}=\frac{e m_\rho^2}{2 \gamma_\rho}=2.52~,~
g_{\Upsilon BB}=g_{\Upsilon B^* B^*}
=\frac{e m_\Upsilon^2}{3 \gamma_\Upsilon}=13.3~.   
\end{eqnarray}
In the above, $\gamma_V$ is the photon-vector-meson mixing amplitude and
can be determined from the vector meson partial decay width to $e^+e^-$.
Also, the light-cone QCD sum rule \cite{pbb} gives  
$g_{\pi BB^*}=10.3$. 
We note that the above values for the coupling constants  
$g_{\pi B B^*}$, $g_{\Upsilon B B}$, and $g_{\Upsilon B^* B^*}$  
deviate appreciably from the SU(5) relation shown in Eq. (\ref{su5r}).  
However, they agree with the predictions from 
the heavy quark symmetries \cite{pbb,hqs,ubb}, i.e., 
\begin{eqnarray}
\frac {g_{\pi B B^*}}{g_{\pi D D^*}} \sim \frac {m_B}{m_D},~
\frac {g_{\Upsilon B B}}{g_{\psi DD}} \sim \sqrt {\frac {m_B}{m_D}}.
\label{hm}
\end{eqnarray}
Since the SU(5) is a broken symmetry, we shall
use in the following calculations the phenomenological values
for these coupling constants.

For the four-point coupling constants, there is no empirical information,
and we thus use the SU(5) relations to determine their values in terms of the
three-point coupling constants, i.e., 
\begin{eqnarray}
g_{\pi \Upsilon BB^*}= g_{\pi BB^*} g_{\Upsilon BB}, ~
g_{\rho \Upsilon B B}= 2~g_{\rho BB} g_{\Upsilon BB}, ~
g_{\rho \Upsilon B^* B^*}=g_{\rho B^* B^*} g_{\Upsilon B^* B^*}.  
\end{eqnarray}

To take into account the composite nature of hadrons, 
form factors need to be introduced at interaction vertices. 
Unfortunately, there are little empirical information on form factors 
involving bottom mesons or $\Upsilon$ states. 
We thus take the form factors to have the usual mono-pole form 
at the three-point $t$ channel and $u$ channel vertices, i.e., 
\begin{eqnarray}
f_3=\frac {\Lambda^2}{\Lambda^2+{\bf q}^2},
\end{eqnarray}
where $\Lambda$ is a cutoff parameter, and 
${\bf q}^2$ is the squared three momentum transfer in the center-of-mass 
frame, given by $({\bf p_1}-{\bf p_3})_{\rm cm}^2$ 
and $({\bf p_1}-{\bf p_4})_{\rm cm}^2$ 
for $t$ and $u$ channel processes, respectively. 
For simplicity, 
we use the same value for all cutoff parameters, 
and choose $\Lambda$ as either $1$ or $2$ GeV 
to study the uncertainties due to form factors.
We also assume that the form factor at four-point vertices has the
following form:
\begin{eqnarray}
f_4=\left ( \frac {\Lambda^2}{\Lambda^2+\bar {{\bf q}^2} } \right )^2, 
\end{eqnarray}
where $\bar {{\bf q}^2}=p_{i,\rm cm}^2+p_{f,\rm cm}^2$ is the average 
value of the squared three momentum transfers in $t$ and $u$ channels.   

\section{Numerical Results}
\label{sec_num}

\begin{figure}[ht]
\centerline{\epsfig{file=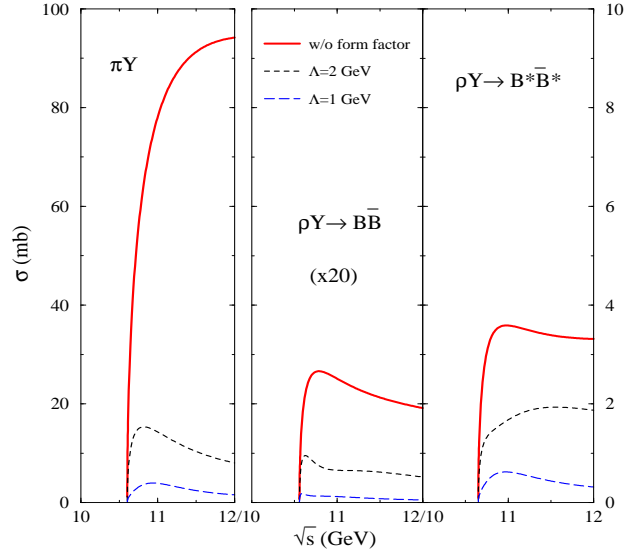,height=3in,width=3in,angle=0}}
\vspace{1cm}
\caption{
Cross sections of $\Upsilon$ absorption 
as a function of the center-of-mass energy of initial-state mesons
with and without form factors. 
The cross section for the process $\rho \Upsilon \rightarrow B \bar B$ 
has been multiplied by a factor of $20$.
}
\label{sigma}
\end{figure}

\begin{figure}[ht]
\centerline{\epsfig{file=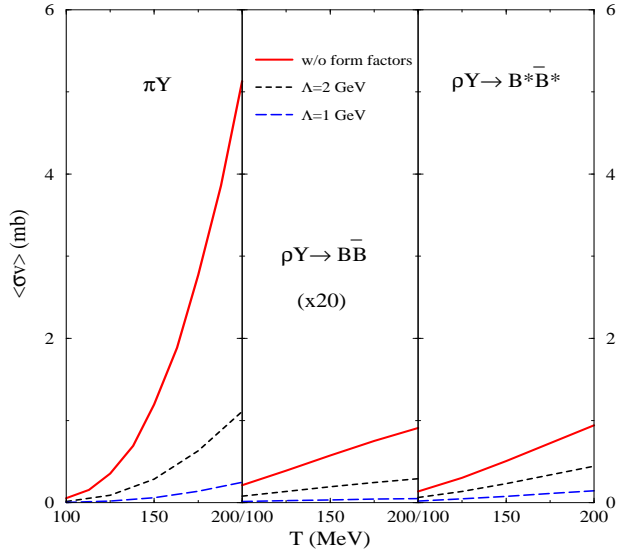,height=3in,width=3in,angle=0}}
\vspace{1cm}
\caption{
Thermal averages of the cross sections for $\Upsilon$ absorption 
as a function of temperature T with and without form factors. 
}
\label{sv}
\end{figure}

In Fig.~\ref{sigma}, we show the cross sections for 
$\Upsilon$ absorption by 
$\pi$ and $\rho$ mesons as a function of 
the center-of-mass energy $\sqrt s$ of the initial-state mesons. 
The cross section for the $\pi \Upsilon$ process 
includes contributions from both $\pi\Upsilon \to B^*{\bar B}$ and 
$\pi \Upsilon \rightarrow B \bar {B^*}$, which have the same cross sections.
Since the centroid value of the $\rho$ meson mass (770 MeV) is used in the
calculation, all processes are endothermic. 
As a result, all cross sections have similar energy dependence near the
threshold. Form factors are seen to strongly suppress the 
cross sections and thus cause large uncertainties in their values.
With the cutoff parameter between $1$ and $2$ GeV, 
the values for $\sigma_{\pi \Upsilon}$ and $\sigma_{\rho \Upsilon}$ 
are roughly $8$ mb and $1$ mb, respectively, 

The thermal average of these cross sections, given by 
\begin{eqnarray} 
\langle \sigma v \rangle 
&=&\frac{\int_{z_0}^{\infty} dz \left [z^2-(\alpha_1+\alpha_2)^2
\right ] \left [z^2-(\alpha_1-\alpha_2)^2 \right ] K_1(z) ~
\sigma (s=z^2{\rm T}^2)}
{4 \alpha_1^2 K_2(\alpha_1)\alpha_2^2 K_2(\alpha_2)} ~,  
\end{eqnarray} 
with and without form factors are shown in Fig.~\ref{sv}. 
In the above, $\alpha_i=m_i/{\rm T}$ ($i=1$ to $4$), 
$z_0={\rm max}(\alpha_1+\alpha_2,\alpha_3+\alpha_4)$, 
$K_n$'s are modified Bessel functions, and 
$v$ is the relative velocity of initial-state 
particles in their collinear frame.  
We note that at a temperature of $150$ MeV, for example, 
both $\langle \sigma_{\pi \Upsilon}v \rangle$ and 
$\langle \sigma_{\rho \Upsilon}v \rangle$ 
are only about $0.2$ mb after including the form factors.
This indicates that $\Upsilon$ absorption by hadronic 
comovers in the final state of high energy heavy ion collisions   
is not expected to be important.

\section{Comparison with $J/\psi$ Absorption by hadrons}
\label{sec_j}

It is interesting to compare our results for the $\Upsilon$ absorption with 
the $J/\psi$ absorption cross sections calculated from a similar 
Lagrangian \cite{jpsih}.  
Rescaling all momenta by the heavy meson mass $m_H$ 
and neglecting all light meson masses, 
the cross sections in Eqs. (\ref{jpion})-(\ref{m3}) without form factors 
can be factorized. Using the scaling relation of Eq. (\ref{hm}) for 
the coupling constants, we obtain
\begin{eqnarray}
\sigma \left (\frac{\sqrt s}{m_H} \right ) 
\propto \frac {g_1^2 g_2^2}{m_H^2} \propto m_H 
~{\rm and}~ \frac {1}{m_H}
\label{scale}
\end{eqnarray}
for the heavy quarkonium scattering cross sections with $\pi$ 
and $\rho$ mesons, respectively. 
From the numerical results shown in Fig. \ref{sigma} 
and Ref. \cite{jpsih}, we find that 
$\sigma_\Upsilon(\sqrt s=14~{\rm GeV})/\sigma_\psi (\sqrt s=5~{\rm GeV}) 
\sim 3, ~0.5, ~{\rm and}~0.4$ 
for the three processes shown in Fig.~\ref{diagrams}, respectively, and these 
ratios agree with the scaling relation of Eq. (\ref{scale}) within 50\%.  

As shown previously, the thermal averages of the 
$\Upsilon$ absorption cross sections by pion and rho mesons 
are both about $0.2$ mb at $T=150$ MeV, which are roughly a factor of
5 to 10 smaller than the thermal averages of $\sigma_{\pi \psi}$ 
and $\sigma_{\rho \psi}$ at the same temperature 
and with the same form factors \cite{jpsih}.
This is mainly due to the larger kinematic thresholds 
(i.e., $m_3+m_4-m_1-m_2$) for the $\Upsilon$ absorption. 
With $m_\rho=770$ MeV, they are
$1.01, 0.33$, and $0.42$ GeV, respectively, 
for the three processes shown in Fig.\ref{diagrams},
compared to $0.64, -0.14$, and $0.15$ GeV for the 
corresponding $J/\psi$ absorption processes.  
The larger threshold for $\Upsilon$ absorption by hadrons
not only prevents more light mesons from participating in the absorption 
process but also causes a larger reduction of the cross sections 
due to the form factors at interaction vertices. 

\section{Discussions and Summary}
\label{sec_sum}

In our study, the hadronic Lagrangian shown
in Eq.~(\ref{lagn2}) is generated from the SU(5) flavor symmetry. 
The resulting PPV, VVV, PPVV and VVVV interaction Lagrangians
are exactly the same as those in the chiral Lagrangian approach \cite{song}. 
However, the SU(5) flavor symmetry is badly broken by quark masses, especially 
by charm and bottom quark masses.  
Although in our study we have used the coupling constants 
determined from either the vector meson dominance model or the QCD sum rules, 
other symmetry-breaking effects are possible and need to be further studied. 
There are also large uncertainties on the values of the 
coupling constants.  
The coupling constant $g_{\pi B B^*}$ given by the QCD sum rules 
\cite{pbb,pbb2} can differ by about a factor of 2, 
and the result from the lattice QCD studies \cite{lattice} is also 
inconclusive due to the large error bar. 
To include the symmetry breaking effects in hadronic models,  
an alternative approach \cite{wise} based on both the chiral symmetry for 
light flavors and the heavy quark spin symmetry for charm and bottom 
flavors may be useful. 

Since there is little experimental information available for
form factors involving bottom mesons,  
significant uncertainties thus exist in results  based on 
hadronic Lagrangians. To reduce these uncertainties, 
studies of $B$ meson decays will be useful. 
For example, the form factor for $B$ meson semileptonic decays \cite{pbb,pbb3},
$B \rightarrow \pi \l \bar {\nu_\l}$, may be related to the form factor for 
the $\pi B B^*$ vertex. Recent studies based on QCD sum rules
\cite{ffq2} have shown that 
the $\pi B B^*$ and $\pi D D^*$ form factors as a function of the 
pion momentum roughly correspond to 
cutoff parameters between $1$ and $2$ GeV if they are
fitted with the mono-pole form. 
Also, the form factor at the $\Upsilon B B$ vertex 
is related to the form factors for decays such as 
$\bar B \rightarrow D \l \bar {\nu_\l}$ \cite{ubb,ubb1,ubb2}.

We note that the absorption cross sections of the $\Upsilon$ by 
heavier mesons such as the kaon and charmed meson can be similarly calculated 
in our model. We have not included them in the present study 
as transport models have shown that the numbers of heavier mesons 
are much less than those of pions and rho mesons in high energy heavy 
ion collisions \cite{urqmd,ampt}. 

We have only considered the absorption of the $\Upsilon (1S)$ 
in hadronic matter. There are also heavier bound states of $b\bar b$,
such as $\Upsilon (2S), \Upsilon (3S), \chi_{bi} (1P)$ and
$\chi_{bi} (2P)$ ($i=0,~1$, and $2$), which can decay
into the $\Upsilon (1S)$. In $pp$ collisions, 
almost half of the final $\Upsilon (1S)$ yield is from the decay of 
these heavier particles \cite{gunion}. To use $\Upsilon (1S)$
as a signal for the quark-gluon plasma in heavy ion collisions thus
requires also information on the absorption cross sections of these 
particles by hadrons. Since they are less bound than 
$\Upsilon (1S)$, these heavy particles are more likely to be dissociated 
in both the partonic \cite{karsch} and hadronic matter \cite{gunion}. 
It will be useful to extend our meson-exchange model to include these heavier 
$b\bar b$ bound states in order to calculate their absorption cross sections. 
 
In summary, we have studied the $\Upsilon$ absorption cross sections 
by $\pi$ and $\rho$ mesons using a hadronic Lagrangian based on 
the SU(5) flavor symmetry. 
Including form factors with a cutoff parameter
of $1$ or $2$ GeV at the interaction
vertices, we find that the values for 
$\sigma_{\pi \Upsilon}$ and $\sigma_{\rho \Upsilon}$ 
are about $8$ mb and $1$ mb, respectively. 
However, due to the large kinematic threshold,   
their thermal averages at a temperature of $150$ MeV 
are both only about $0.2$ mb. 
Our results thus suggest that the absorption of directly produced $\Upsilon$ 
by comoving hadrons is unlikely to be important in high energy 
heavy ion collisions. 

\section*{Acknowledgments} 

This work was supported in part by the National Science Foundation under 
Grant No. PHY-9870038, the Welch Foundation under Grant No. A-1358,
and the Texas Advanced Research Program under Grant No. FY99-010366-0081.

\pagebreak
{}

\end{document}